%% file: main.tex
\documentclass[submission]{eptcs}
\providecommand{\event}{PxTP 2021} 
\usepackage{breakurl}             
\usepackage{underscore}           
\usepackage[utf8]{inputenc} 
\usepackage[T1]{fontenc}
\usepackage{xcolor}         
\usepackage{listings}
\usepackage{xspace}
\usepackage{alltt}
\usepackage{graphicx}
\usepackage{amssymb}
\usepackage{amsthm}
\usepackage{mathtools, bm}
\usepackage{amssymb, bm}
\usepackage[all]{xy}

\input{macros}

\theoremstyle{definition}
\newtheorem{example}{Example}

\lstdefinelanguage{Coq}{morekeywords={Variable,Inductive,CoInductive,Fixpoint,CoFixpoint,Definition,Lemma,Theorem,Axiom,Local,Save,Grammar,Syntax,Intro,Trivial,Qed,Intros,Symmetry,Simpl,Rewrite,Apply,Elim,Assumption,Left,Cut,Case,Auto,Unfold,Exact,Right,Hypothesis,Pattern,Destruct,Constructor,Defined,Fix,Record,Proof,Induction,Hints,Exists,let,in,Parameter,Split,Red,Reflexivity,Transitivity,if,then,else,Opaque,Transparent,Inversion,Absurd,Generalize,Mutual,Cases,type,of,end,Analyze,AutoRewrite,Functional,Scheme,params,Refine,using,Discriminate,Try,Require,Load,Import,Scope,Set,Open,Section,End,match,with,Ltac,fun,forall,exists,Goal},sensitive,morecomment=[n]{(*}{*)},morestring=[d]",literate={=>}{{$\Rightarrow$}}1 {>->}{{$\rightarrowtail$}}2{->}{{$\to\,$}}1{\/\\}{{$\wedge$}}1{|-}{{$\vdash$}}1{\\\/}{{$\vee$}}1{~}{{$\sim$}}1{⟦}{{$\llbracket$}}1{⟧}{{$\rrbracket$}}1}[keywords,comments,strings]
\title{General Automation in \Coq through Modular Transformations}
\author{Valentin Blot
\institute{LMF, Inria, Université Paris-Saclay\footnote{This work is funded by a Nomadic Labs-Inria collaboration.}}
\email{Valentin.Blot@inria.fr}
\and
Louise Dubois de Prisque
\institute{LMF, Inria, Université Paris-Saclay*}
\email{Louise.Dubois-de-Prisque@inria.fr}
\and
Chantal Keller
\institute{LMF, Université Paris-Saclay*}
\email{Chantal.Keller@lri.fr}
\and
Pierre Vial
\institute{LMF, Inria, Université Paris-Saclay*}
\email{Pierre.Vial@inria.fr}
}
\def\titlerunning{Automation through modular transformations}
\def\authorrunning{V. Blot, L. Dubois de Prisque, C. Keller, P. Vial}

\begin{document}
\sloppy
\maketitle

\begin{abstract}
Whereas proof assistants based on Higher-Order Logic benefit from external solvers' automation, those based on Type Theory resist automation
and thus require more expertise. 
Indeed, the latter use a more expressive logic which is further away from first-order logic, the logic of most automatic theorem provers.
In this article, we develop a methodology to transform a subset of \Coq goals into first-order statements that can be automatically discharged 
by automatic provers.
The general idea is to write modular, pairwise independent transformations and combine them.
Each of these eliminates a specific aspect of \Coq logic towards first-order logic.
As a proof of concept, we apply this methodology to a set of simple but
crucial transformations which extend the local context with proven first-order assertions
that make \Coq definitions and algebraic types explicit.
They allow users of \Coq to solve non-trivial goals automatically. 
This methodology paves the way towards the definition and combination of more complex transformations, making \Coq more accessible. 
\end{abstract}

\section{Introduction}
\label{sec:intro}

\ignore{Voici le sens des modifications que je suggère. D'abord, une légère modification du plan:\\
a: problématique de l'automatisation dans \Coq\\
b: un exemple détaillé\\
c: la soution de \SMTCoq\\ 
Ceci me paraît motivé pour ces raisons:\\
1: Dans la mesure où l'exemple du comb est un but du 1er ordre, je pense qu'il vaut mieux mettre en contexte la logique du 1er ordre/des SMT solvers.\\
2: avant de parler du comb, i.e. un exemple où \Coq montre son manque d'automatisation, il faut problématiser d'abord la notion d'interactivité.\\
3: puisqu'on commence à parler de tactique, il faut aussi parler du noyau de \Coq. Actuellement, les explications sur le noyau se trouvent en deux endroits: au début de la section 1 puis en début de la section 2. Dans la mesure où la notion de noyau et des surcouches de \Coq est plus générale que l'exploitation particulière qui en est faite par 
SMTCoq, je pense qu'il faut en parler avant (et séparément) de parler de \SMTCoq.
\\ 
4: un possibilité alternative serait de commencer directement par l'exemple, mais sans parler de tactique, en se bornant juste à remarquer que le théorème comb devrait être trivial à prouver, mais qu'on doit quand même indiquer la preuve de façon assez détaillée.\\
5: dans les deux cas, je pense que le traitement de comb avec \veriT devrait être décrit avec moins de détails techniques.\\
}

\paragraph{Interaction \vs automation.}
The \Coq proof assistant allows us to prove theorems \emph{interactively}: that is, given a \Coq goal $\ttG$ (\ie a statement to prove), the user has to write a \emph{proof} $\ttP$ consisting of inference steps with \textit{ad hoc} premises and conclusion. Then, they  can check whether the given proof is correct by writing \lstinline !Qed!:
the proof is then certified (or rejected) by a \textbf{logical kernel} implemented in OCaml which type-checks the proof-term constructed by the user. In practice, proof assistants support a limited form of \textit{automation}: in \Coq, 
 the user can use keywords called \textit{tactics}, which  
operate logical transformations on a goal and its hypotheses until
solving it, \eg a tactic may try to apply some inference/typing rules as many times as possible and relieve the user of some bureaucratic work. An important idea\label{disc:about-the-kernel} to keep in mind is that tactics may use elaborate tools (including plugins, auxiliary softwares, \textit{etc}\ldots) to produce whole proof terms,  \textit{including flawed ones}. However, eventually, \textbf{any proof term,} 
whether it is written by the user or produced by a tactic or an external tool, \textbf{will be type-checked by the kernel of \Coq}, and this is how the trust in \Coq lies in the one we have in its kernel.

Anyway, automation in \Coq is limited: in most cases (including many trivial proofs),  the user has to provide the bigger part of the structure and the steps of the proof: we will give a simple example just below.

This situation is strikingly different from \emph{Automated Theorem
  Provers}, such as first-order solvers or \textbf{SMT solvers}, which find a
proof without the user having to find the proof steps if the problem is
expressed in a suited way.
However, these automated provers have two main
limitations:
\begin{itemize}
\item One needs to trust their whole code, and not a small kernel.
\item Most of them handle only \textbf{first-order logic (FOL)} whereas \Coq is based on the \textbf{Calculus of Inductive Constructions (CIC)}, which is far richer: (1) CIC is polymorphic,  and thus allows quantifying not only over objects (such as integers) but also on types, \eg in CIC, one may define polymorphic list concatenation \lstinline!++! whose type is \lstinline!forall (A : Type), list A -> list A -> list A!
(2) CIC features dependent typing, so that types may depend on terms and other types, \eg \lstinline!Vec A n! specifies the type of vectors of length \lstinline!n! on the carrier \lstinline!A!, and thus depends on the number \lstinline!n! and the type \lstinline!A! (3) CIC enables higher-order computation.
\end{itemize}

The logical expressiveness of CIC makes it possible to handle complex
aspects of programming languages and to prove properties about programs
in \Coq, in particular fine-grained specifications that an automatic
prover based on first-order logic could not understand. For
instance, dependent typing allows the specification of equalities beyond
decidable datatypes, \eg in \Coq one may directly specify that
two \emph{functions} are equal with an equality \lstinline!f = g!, which
is not possible in general in automatic solvers. Yet, it is precisely
because most automatic provers only tackle FOL that they admit a lot of
powerful automation.

\paragraph{Improving the automation of \Coq within first-order logic.}
To sum up the situation, on the one hand, we have \Coq, which is based
on CIC and is a rich specification language, but for that reason, it is
difficult to automatize its proof search. On the other hand, automatic
provers are based on a more limited logic (FOL) but feature extremely
efficient proof-search heuristics. Moreover, the trust we have in the
former is based on the implementation of a small, isolated kernel. This
can be summarized in Table~\ref{table:oq-vs-SMTsolvers}.
  \begin{table}
  \begin{center}
    \begin{tabular}{|r|l|l|}
  \hline
   & Automatic provers & \Coq \\
  \hline
  Expressivity & First-order logic  & {\bf CIC} \\
  Safety & Trust the whole software & {\bf Kernel for proof checking } \\
  Automation & {\bf Automatic proof} & User-guided proof \\
  \hline
  \end{tabular}\\
  \hspace*{6.8cm} 
  \caption{Pros (bold) and Cons of automatic provers and \Coq}
  \label{table:oq-vs-SMTsolvers}
  \end{center}
  \end{table}

 Such differences are to be expected, since proof assistants and
 automatic provers do not have the same purposes and uses.
  However, as it turns out, in \Coq, even a \textit{simple proof of
    first-order logic}, \eg on decidable datatypes, can be
  tedious and force the user to provide a lot of input, whereas an
  automatic prover would automatically find a proof. Introducing
  more automation in \Coq would thus be very useful to spare the user
  some trivial parts of a proof.

Let us illustrate this with two examples.

\begin{example}[Dealing with datatypes]\label{ex:datatypes}
We use the function
\lstinline!hd_error! from the \lstinline!List! module of \Coq Standard
Library, of type
\lstinline!forall {A:Type}, list A -> option A!
(the curly brackets mean that \lstinline!A! is an implicit argument,
that is to say it can be omitted) defined by
\lstinline!hd_error l = Some x!
when
\lstinline!l = x :: l0!
(\ie \lstinline!l! not empty, \lstinline!x :A!, \lstinline!l0! a list of elements of type \lstinline!A!) and
\lstinline!hd_error [] = None!. We then prove:
\begin{lstlisting}
Goal forall l (a:A), hd_error l = Some a -> l <> nil.
\end{lstlisting}
in a context where \lstinline!A! is a type variable\footnote{This is
  handled in Coq thanks to the section mechanism. It allows the user to introduce section-local variables that can be used in other declarations in the section.}.
A typical \Coq proof is:
\begin{lstlisting}
Proof.
 intros l a H. intro H'. rewrite H' in H. simpl in H. discriminate H.
Qed.
\end{lstlisting}

The statement is straightforward, its proof relies on the fact that two
different constructors always output different values, \eg
\lstinline!Some x! and \lstinline!None! cannot be equal, neither
\lstinline!x :: l! and \lstinline![]!. Yet, in \Coq, the user
(especially a non-expert one) has to be highly precise in the way they
compose their keywords, even though they would not even bother writing the
proof on paper. It may even be more frustrating that this is a statement of
first-order logic and as such, it would automatically  be dealt with by a
first-order prover, provided it knows about (1) the definition of the
function \lstinline!hd_error! and (2) the datatypes \lstinline!list! and
\lstinline!option!.
\end{example}

\begin{example}[Calling lemmas]\label{ex:search-lemma}
  In this example, we recall some of the annoyances met while using lemmas in \Coq.   
  We consider a Boolean search function:
\begin{lstlisting}
Fixpoint search {A : Type} {H: CompDec A} (x : A) l := 
  match l with 
  | [] => false
  | x0 :: l0 => eqb_of_compdec H x x0 || search x l0
end.
\end{lstlisting}

Not going into details, the implicit argument \lstinline!H: CompDec A!
specifies that \lstinline!A! is a \emph{decidable type}, \ie that
equality is decidable on \lstinline!A!. Boolean equality can then be
computed with the function \lstinline!eqb_of_compdec H!\footnote{There
  are multiple ways of representing a decidable equality in \Coq; we use
  the representation from the \SMTCoq plugin since we will use it as a
  back-end (see later).}.

By induction, we prove
\lstinline!search_app : forall {A: Type} {H : CompDec A} (x: A) (l1 l2: list A), search x (l1 ++ l2) = (search x l1) || (search x l2)!.
Now, let us consider a typical \Coq proof of the following simple statement:

\begin{lstlisting}
Lemma search_lemma : forall (x: Z) (l1 l2 l3: list Z),
  search x (l1 ++ l2 ++ l3) = search x (l3 ++ l2 ++ l1).
Proof.
  intros x l1 l2 l3. rewrite !search_app.
  rewrite orb_comm with (b1 := search x l3).
  rewrite orb_comm  with (b1 := search x l2) (b2 := search x l1).
  rewrite orb_assoc. reflexivity.
Qed.
\end{lstlisting}
As expected, the proof uses the commutativity and the associativity of
Boolean disjunction \lstinline!||!, which can be found in the module
\lstinline!Coq.Bool.Bool! of the Standard Library. We start by using the
lemma \lstinline!search_app! 4 times to eliminate \lstinline!++! from
the statement, which can be done automatically with the \lstinline!!!
operator that rewrites as much as needed.
However, the order of commutativity/associativity uses must be carefully
chosen. Moreover, the instance of the bound variables of the
commutativity lemma \lstinline!forall b1 b2 : bool, b1 || b2 = b2 || b1!
must be specified: the same proof without specifying them
fails, since \Coq would try to rewrite only the leftmost-outermost
\lstinline!||!, which would not work. Actually, although the user
recognizes the proof as trivial, they must keep a keen eye at each
step on the local proof context to determine which lemma must be used
with which instances, and sometimes, they have to print lemmas to
identify the names of bound variables they need to instantiate. All this
may appear as a nuisance compared to writing a formal proof with a pencil
and a paper and deter new users of \Coq.
\end{example}

\paragraph{Our contribution: linking first-order \Coq goals with
  automated provers}
An interesting observation about Examples~\ref{ex:datatypes} and
~\ref{ex:search-lemma} is that their statements and proofs
pertain to first-order logic (with decidable equalities). As such, they
\emph{should} be automatized.

In this paper, we provide a methodology to reconcile \Coq goals with the
logic of first-order provers. This methodology consists in
\begin{enumerate}
\item implementing pairwise independent logical transformations: each
  transformation encodes one aspect of \Coq logic as formulas in a less expressive logic (until reaching first-order formulas in \Coq) 
  and establishes a
  soundness proof of this encoding;
\item providing strategies to combine these transformations in order to
  automatically translate \Coq goals into fully explicit first-order
  logic goals.
\end{enumerate}
We also implement this methodology as a new \Coq tactic called \snipe
(for the bird known in french as \textit{bécassine des
  marais}) so that the proofs of the two lemmas presented above become
only one single call to this tactic, passing the required lemma
\lstinline!search_app! in the second case.

The tactic \snipe is two-fold.
\begin{enumerate}
\item As presented before, we implemented five small logical
  transformations, and a strategy that combines them. The
  transformations presented in this paper deal with definitions,
  datatypes and polymorphism, and thus the strategy transforms a \Coq
  goal containing these features into a fully first-order goal.
\item Then, we use the SMT solver \veriT as a back-end to discharge the
  obtained first-order goal, available through the \SMTCoq\footnote{\SMTCoq is
  available at
  \url{https://smtcoq.github.io}.} plugin\cite{DBLP:conf/cav/EkiciMTKKRB17},
  which enables safe communication between \Coq and SMT solvers.
\end{enumerate}

In step 2, we benefit from the automation provided by the SMT solver
\veriT, which is able in particular to perform Boolean computation (as
in Example~\ref{ex:datatypes}), Linear Integer Arithmetic (LIA)
or relieve the user of the burden of finding the right instantiations of
the lemmas. An important observation is that in step 2, the use of
\SMTCoq could be replaced by any tactic solving first-order logic.

The remainder of the paper is dedicated to explaining the concept of
\snipe in details. In the next section, we explain our methodology. In
\Sec~\ref{sec:transfos}, we present examples of useful
logical transformations together with their implementations. In
\Sec~\ref{sec:poc}, we explain the full \snipe tactic, as well as more
examples of its power. We finally present the state of the art before
concluding.

The source code can be found at
\url{https://github.com/smtcoq/sniper/releases/tag/pxtp21}.

\section{From the logic of \Coq to first-order logic: modular transformations}
\label{sec:methodology}

\paragraph{From the Calculs of Inductive Constructions to first-order logic}

As we saw in the introductory examples, \Coq, which is based on CIC, is mostly not
automatized even for simple proofs: while it now enjoys very efficient
decision procedures for dedicated
theories~\cite{DBLP:conf/types/Besson06,DBLP:conf/tphol/GregoireM05},
attempts for general automation has not truly succeeded yet (see
\Sec~\ref{sec:soa} for a detailed comparison).

Our contribution is to propose a new approach for general automation,
  based on (1) small transformations of a CIC goal towards a goal of first-order logic (2) stating and proving first-order properties in the local context of a \Coq proof. 
  Then, the first-order goal and the associated first-order properties can be sent to any external automated prover based on first-order logic. 
  As we will see in the next paragraph, the user may choose which transformation they apply.

We choose FOL as our target logic because
a lot of work has been done to automatize reasoning in this logic. So, after
calling our transformations, the user can choose any way to
automatically prove a first-order goal: \eg an automatic prover
certified in \Coq, a tool calling external solvers, tactics like
\lstinline!firstorder! or
\lstinline!crush!\cite{DBLP:books/daglib/0035083}. In
\Sec~\ref{sec:poc}, we provide a fully-automated tactic
that applies the transformations presented in this paper and solves the
resulting first-order goal using the \SMTCoq plugin.



\paragraph{Modular and independent transformations}
Formally, a logical transformation from the language of \Coq 
 to itself is a function $f$ from the terms and formulas of \Coq to
 themselves. Leaving aside the details on the nature of the function $f$
 for now, we are interested in \textbf{sound transformations}, \ie
 transformations $f$ such that, given any \Coq statement $G$ in the
 domain of $f$, we have $f(G) \Rightarrow G$. This means that it is
 enough to prove $f(G)$ for $G$ to be valid. More precisely, we are
 interested with sound transformations $f$ such that $f(G)$ is a
 first-order formula: indeed, if a \Coq goal $G$ is left to prove, we
 may transform $G$ into $f(G)$ and then send $f(G)$ to an automated
 theorem prover dealing with first-order. If this succeeds, it means
 that $G$ is valid.

We are actually also interested in functions $g$ from a subset of \Coq
formulas such that, given a \Coq statement $G$, $g(G)$ outputs (a list
of) \emph{valid} first-order logic statements in \Coq that may help
proving $G$. We also call such a function $g$ (which produce
auxiliary first-order statements) a sound transformation.

Our approach is to develop modular and independent transformations: each
of them encodes one important aspect of CIC. 
The aim is to tackle only one aspect at a time: it facilitates the proof
of soundness whenever we need to write one in \Coq.
Indeed, as the transformations are simple, they preserve as much as
possible the structure of the source formula.
We expect that they are easier to implement than a bigger encoding. 
In addition, this methodology facilitates the debugging and 
allows us to know precisely which fragment of \Coq we can handle.
The transformations can be composed in different ways: we can use them separately,
combine them all, or only some of them, either by using a default tactic
provided for a user who does not want to think about
which method they should choose, or by writing them one by one. 
Moreover, they are independent from the technology used for first-order
proving in the end. This is illustrated by the left part of Figure~\ref{fig:snipe}.


\paragraph{Certifying and certified transformations}


There are two ways for writing logical transformations $f$ from \Coq to itself. As we will see, both need to resort to the \textbf{meta-language} of \Coq at some point: they feature functions which cannot be defined in the core language of \Coq, but only in extra-layers outside its kernel. However, they do not work the same way from the logical point of view.
\begin{enumerate} 
\item \textbf{Certified transformations.} The function $f$ may be a \Coq
  function. This relies on an internal representation of \Coq terms
  inside \Coq, that we call \lstinline!term! (see
  Example~\ref{ex:reif-example}). It comes with two meta-language
  transformations: the {\bf reification} takes a \Coq term as a
  parameter and outputs its reification (in the type \lstinline!term!)
  and the {\bf dereification} is the converse. In this situation, $f$ is
  a \Coq function (\emph{not} a meta-function) of type
  \lstinline!term -> term! and we may prove that, for all \lstinline!t!
  of type \lstinline!term!, $f$ \lstinline!t! implies \lstinline!t! (up
  to some implicit reification). We call this kind of transformation a
  \emph{certified} one, because the soundness is proved \emph{once and
    for all}, \emph{in a \Coq statement}.
\item \textbf{Certifying transformations.} The function $f$ may be a meta-language function, \ie $f$ is not a \Coq function. 
In that case, it is \emph{not possible} to write a \Coq statement specifying that, for instance, for all $G$ of type \lstinline!set!, $f(G)$ implies $G$. 
However, we may write another meta-language function $g$ such that, for all $G$, $g(G)$ generates on the fly a proof of $f(G) \rightarrow G$.  
Such a transformation is said to be \emph{certifying}, because its soundness is not previously established. 
It takes a local context and a \emph{specific} goal, and it operates the transformation, which will be type-checked at the end by 
the kernel (when \lstinline !Qed! is written) \emph{every time we use it}. 
\end{enumerate}

\begin{figure}[!ht]
\begin{center}
\includegraphics[scale=0.55]{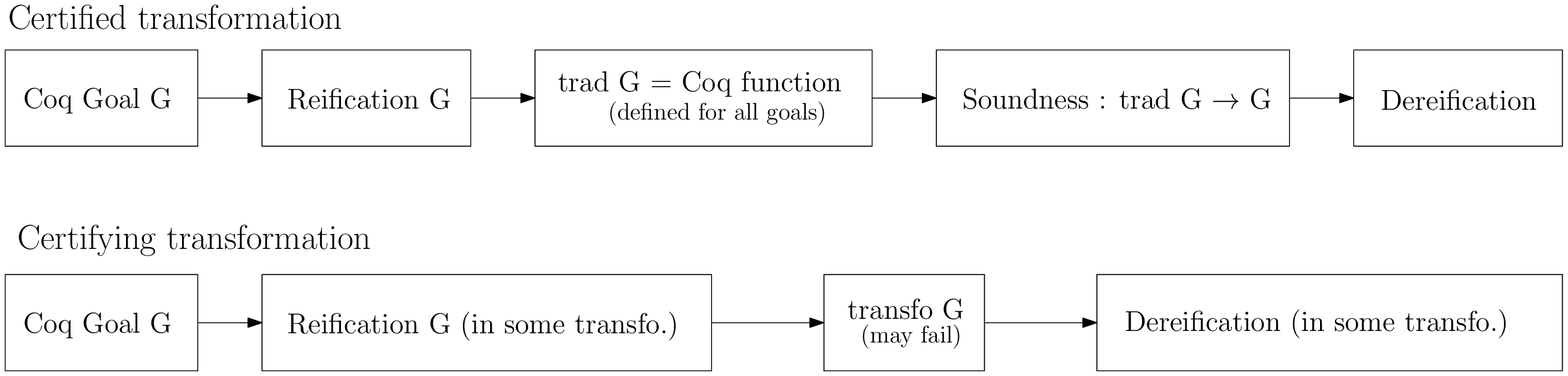}
\caption{Difference between certified and certifying transformations}
\label{fig:certif}
\end{center}
\end{figure}

The differences between certified and certifying transformations are 
summarized in Figure~\ref{fig:certif}.
In the work presented in this article, we follow the paradigm of
certifying transformations. Indeed, the former require we work only with the reified syntax of the terms of CIC and 
even for a simple transformation, the proof of soundess is hard (thousands lines of code). 
But in some situations, it could be useful to use this solution, because it covers all cases. 


\paragraph{Metalanguage tools}

Most logical transformations analyze the syntax of terms in the source language and produce new statements. 
But in \Coq, we do not directly have access to the syntax of terms. 
This is the reason why we need meta-programming tools
as we saw in the previous paragraph.
Many different
tools are available now (see \Sec~\ref{sec:soa}), and we used two of them:
\Ltac~\cite{DBLP:conf/lpar/Delahaye00} and
\MetaCoq~\cite{DBLP:journals/jar/SozeauABCFKMTW20}, which offer
different advantages.

\begin{itemize}

\item \Ltac is a tactic language in Coq which is based on pattern matching, recursion and backtracking. 
It allows its users to match on the syntax of Coq terms and also on the local context, for building new goals and terms.
It is also possible to execute basic tactics which operate on kernel terms, called tacticals, in a \Ltac function. 
\Ltac cannot be seen as a proper meta-programming language as there is no data type which contains the reified syntax of Coq terms. 
In our transformations, we use \Ltac whenever we need to use a particular function on every hypotheses of the context.
Once this function is applied, its result is certified in \Ltac by a combination of simple Coq tacticals. 
Sometimes, it is also useful when a superficial access to the syntax of a Coq term is required, but \Ltac is less convenient for analyzing the syntax completely.

\item \MetaCoq enables a more fine-grained analysis on \Coq terms. 
Indeed, this plugin includes an inductive type \lstinline !term! which corresponds exactly to the \Coq counterpart of \OCaml kernel terms,
and comes with the reification and dereification transformations described above.
As it offers an easy access to the syntax of Coq terms, our transformations often use \MetaCoq.
The problem with \MetaCoq terms is that the reified syntax is not very readable: the variables are represented by de Brujin indexes, and
there is no notation to reduce the size of the terms. 
This is the reason why we prefer \Ltac to get information about the initial goal, and \MetaCoq when we need to build a new term. Such a flexible approach is possible, 
since we build certifying (not certified) transformations.
\end{itemize}

\begin{example}[Reification]\label{ex:reif-example}
The \MetaCoq representation of the \Coq term
\lstinline!forall (A:Type), A -> A!
is
\lstinline!tProd (name "A") (tSort type_reif) (tProd unnamed (tRel 0) (tRel 1))!.
The constructor \lstinline!tProd! corresponds to the \Coq \lstinline!forall! dependent
product binder (and the $\rightarrow$ is simply a notation
for a non-dependent product). The type of the variable in this product is 
\lstinline!type_reif! (not going into details, this is 
the \MetaCoq reification of \lstinline!Type!). \lstinline!tRel 0! and \lstinline!tRel 1! are the variables, represented
by their De Brujin indexes.
\end{example}

\section{Examples of transformations}
\label{sec:transfos}

We describe now some of the transformations we have implemented.
Most of them are motivated by the following facts.
\begin{enumerate}
\item When \Coq and an external automated theorem prover communicate,
  \textbf{a lot of symbols defined in \Coq will not be intrepreted}. For
  instance, the function \lstinline!hd_error! used in
  Example~\ref{ex:datatypes} may be uninterpreted in the external
  prover, which does not know anything about it except its type. In
  general, pattern-matching may be uninterpreted. Moreover, algebraic
  data types and their constructors may also be left uninterpreted. For
  instance, an external prover may not know that lists have basic
  properties, \eg
  \lstinline!x1 :: l1 = x2 :: l2!
  implies
  \lstinline!x1 = x2! and \lstinline!l1 = l2!, or that
  \lstinline![] <> x :: l!,
  whereas these two \emph{first-order statements} just come from the
  definition of the type \lstinline!list!.
\item \textbf{\Coq handles higher-order objects (in particular,
equalities about such objects), whereas first-order provers do not}. For
instance, \lstinline! fun x => (x + 1) ** 2 = fun x => x ** 2 + 2 x + 1!
is an equality between functions (of type \lstinline!Z -> Z!) and cannot
be directly interpreted in first-order logic. Yet, it may be sufficient
to consider its first-order consequence
\lstinline!forall x, (x + 1) ** 2 = x ** 2 + 2 * x  + 1!, which is a
quantified equality on the type \lstinline!Z!.
\end{enumerate}
Such transformations, while simple at first glance, are already
mandatory to give a bridge between a \Coq goal and a first-order prover:
the first kind of transformation makes the global context explicit,
whereas the second kind encodes higher-order aspects.

As explained in the previous section, we implemented these encodings as
certifying transformations (cf. Figure\ref{fig:certif}), using meta-programming. 
Transformations \ref{sec:transfos:definitions} and \ref{sec:transfos:monomorphization} do not need reification 
while others do.
The approach of each
transformation is the following: by scanning the goal, it states and
proves \emph{in \Coq} various auxiliary lemmas of first-order logic
about the terms encountered in the goal. 
The proofs are done easily by applying \Coq basic tactics. 
These lemmas are stored in the local
context. It may also perform some transformations on the goal, so that it
becomes first-order.



\subsection{Definitions}
\label{sec:transfos:definitions}

The first transformation\footnote{See file \lstinline !definitions.v!.}
makes user-defined terms explicit. For instance, on
Example~\ref{ex:datatypes}, it will add the following assertion to the
local context:
\begin{lstlisting}
hd_error_def : hd_error =
  (fun (A : Type) (l : list A) =>
    match l with | [] => None | x :: _ => Some x end)
\end{lstlisting}
This assertion is not first-order yet (it will be transformed again by
the next two transformations) but it allows one to have access to the
definition of the constant.

For its implementation, it mainly uses the \Coq tactic \lstinline!unfold!, which takes an
identifier (the name of a previously defined term) as a parameter and replaces it by 
its definition.
This tactic, \lstinline
!get_def!, takes the name of a constant and adds its definition as a
proven hypothesis. The example right above is obtained by applying this
tactic on the identifier \lstinline!hd_error!.
While \lstinline !unfold! may cause the goal or the hypotheses to become verbose and not particularly understandable for the user, 
\lstinline !get_def! will keep the definitions in separated statements.
In the tactic which combines all of our transformations, we apply \lstinline!get_def! recursively to 
all the definitions that occur in the hypotheses and in the goal. 


\subsection{Expansion}
\label{sec:transfos:expansion}

Note that in the example for the tactic \lstinline!get_def!, the added hypothesis pertains to a higher-order function (because of \lstinline!fun (A : Type)...!).
A good way to make the equality deal with first-order objects is to 
apply the functional definition to an arbitrary argument of its 
domain\footnote{See file \lstinline !expand.v!.}. 
That is, instead of writing $f = \lambda x. \: t(x)$, we write: $\forall x, f(x) = t(x)$ where $x$ is a fresh variable, and $t$ the unfolded definition of $f$. 
We did not use a transformation which encodes partial application 
with an applicative symbol because it may lead to bigger terms and it is not necessary here. We have left it for future work and more complex cases.
The tactic takes a hypothesis \lstinline!H! of the form $t=u$ where $t \: : \: A_{1} \rightarrow \ldots \rightarrow A_{n} \rightarrow B$, and asserts and proves automatically the hypothesis 

$\forall x_{1} \: : \: A_{1}, \:  \ldots, \forall x_{n} \: : \: A_{n}, \ t (x_{1}, \: \ldots, \: x_{n}) = u (x_{1}, \: \ldots, \: x_{n})$. 

If we come back to our running example, once we get the axiom \lstinline
!hd_error_def! in our context, we can apply
\lstinline !expand hd_error_def! to obtain a new assertion in the local
context:
\begin{lstlisting}
H0 : forall (A : Type) (l : list A), hd_error l =
  match l with | [] => None | x :: _ => Some x end
\end{lstlisting}

The proof of this statement is really simple in \Coq. 
It suffices to use the tactic
\lstinline !rewrite H!, followed by \lstinline!reflexivity!, which
checks that both members of the equality are convertible in CIC.
But the construction of the statement is harder. 
First, the tactic takes a hypothesis \lstinline!H! (\lstinline!hd_error_def! in our case) and finds, thanks to the \Ltac\ value-function \lstinline!type of!, its type
$T$, which must be an equality $t=u$.
In order to have an easy access to the domain and codomain of $t$ and $u$, we reify $T$. 
That is, we call a tactic \lstinline !quote_term! from the \MetaCoq plugin, which takes a \Coq term and returns its reified syntax.
Then we can perform syntactic operations on this term by implementing auxiliary functions directly in \Coq.
Our tactic is thus divided in four parts: 
\begin{itemize}
\item The main tactic \lstinline!expand! reifies the hypothesis \lstinline!H! and calls the auxiliary functions.
\item The first auxiliary function \lstinline!list_of_args_and_codomain! finds the common type of $t$ and $u$: $A_{0} \rightarrow \ldots \rightarrow A_{n} \rightarrow B$ and returns the pair 
$([A_{0}; \ldots ; A_{n}], B)$. 
All the terms involved here are \MetaCoq\ terms. 
\item The second auxiliary function \lstinline!gen_eq! constructs the reified equality that we want. It is defined recursively on the list $[A_{0}; \ldots ; A_{n}]$ by the following equations. 

$\text{\lstinline!gen_eq!} \: ([], B, t, u) \triangleq \: t=_{B}u$

$\text{\lstinline!gen_eq!} \: ([A_{0}; \ldots ; A_{n}],\: B,\: t,\: u) \triangleq \forall x_{0}\: : \: A_{0}, \: \text{\lstinline!gen_eq!} ([A_{1}; \ldots ; A_{n}], B, t \; x_{0}, \: u \; x_{0})$
The \lstinline!gen_eq! function deals with De Brujin indices by lifting these at every recursive call.
\item The generated equality is then unquoted and proved by using \Coq tactics.
\end{itemize}


\subsection{Elimination of fixpoints}

For the sake of simplicity, we have so far taken the example of a
non-recursive function (\lstinline!hd_error!). However, it is very often
the case that functions are defined recursively on datatypes. In this
section, we take the example of the recursive function
\lstinline!length! computing the number of elements in a list. If we
expand its definition as presented in the previous sections, we get:
\begin{lstlisting}
H : forall (A : Type) (l : list A), length l =
  (fix length_anon (l : list A) : nat :=
    match l with | [] => 0 | _ :: l' => S (length_anon l') end) l
\end{lstlisting}
As \lstinline!length! is a fixpoint, its definition is hidden in an
anonymous function denoted by \lstinline!fix!. Again, we transform this
expression to make it more intelligible by automatic provers.
This is the role of the tactic \lstinline!eliminate_fix!\footnote{See file \lstinline!elimination_fixpoints.v!} 
that replaces the anonymous function with the constant it defines. 
Here, \lstinline!eliminate_fix H! asserts and proves the new hypothesis \lstinline!H0!:
\begin{lstlisting}
H0 : forall (A : Type) (l : list A), length l =
  match l with
    | [] => 0
    | _ :: l' => S (length l')
  end
\end{lstlisting}
Thanks to this tactic, we now have access to the body of the
definition of \lstinline!length! as for non-recursive functions.

\subsection{Elimination of pattern matching}
\label{sec:transfos:pattern}

Definitions by pattern matching, such as the examples of the
previous two
subsections, are not understandable for automated provers.
More generally, they are not part of the syntax of first-order terms. 
Thus, instead of getting a function defined by pattern matching, we would like to have one statement for each pattern.
We implemented a tactic\footnote{See file \lstinline!elimination_pattern_matching.v!.} which does precisely this. 
In order to present it, let us continue our example from~\ref{sec:transfos:expansion}. 
The tactic \lstinline!eliminate_pattern_matching H0! (where \lstinline!H0! is the hypothesis generated in~\ref{sec:transfos:expansion}), produces two hypotheses:
\begin{lstlisting}
H1 : forall (A : Type), hd_error [] = None
H2 : forall (A : Type) (x : A) (l : list A), hd_error (x::l) = Some x
\end{lstlisting}


The tactic works in five steps. Note that the formula to which it is applied must be of the form
\lstinline !forall (x$_0$: A$_0$) ... (x$_i$: A$_i$), E[match x$_i$ with ...]! where 
\lstinline !E! is an environement with (possibly) free variables. 
\begin{itemize}
\item The index $i$ of the matched variable is computed with a combination
of a dummy subgoal and a metavariable allowing to pass on the result to the main goal.

\item The formula is reified and the reified
types $A_{0}, \ldots ,A_{i}$ are retrieved. 

\item An independent tactic scans the global environment in which the inductive definition of $A_{i}$ can be found. It returns the list of its reified constructors 
$[C_{0}; \ldots ; C_{j}]$ and their reified types $[T_{0,0} \rightarrow \ldots \rightarrow T_{0,n_{0}}\rightarrow A_{i}; \ldots; T_{j,0} \rightarrow \ldots \rightarrow T_{j,n_{j}}\rightarrow A_{i}]$.
\item For each constructor $C_{k}$, we construct the following statement:

\begin{lstlisting}
forall (x$_0$: A$_0$) ... (x$_{i-1}$: A$_{i-1}$),
  forall (a$_{k,0}$: T$_{k,0}$) ... (a$_{k,n_k}$ : T$_{k,n_k}$),
  E(match C$_k$ a$_{k,0}$ ... a$_{k,n_k}$ with ...)
\end{lstlisting}
\item Each statement is unquoted, asserted and proved by \lstinline
  !intros; rewrite H0; reflexivity!.
\end{itemize}

\subsection{Monomorphization}
\label{sec:transfos:monomorphization}

Most automatic provers do not support polymorphism. 
In other words, they cannot prove lemmas about functions that can be
defined on any type of data.
Typically, as we saw in Example~\ref{ex:search-lemma}, the lemma
\lstinline!search_app! is polymorphic, and thus cannot be sent directly
to most provers. However, only its instance on \lstinline!Z! is useful for proving \lstinline!search_lemma!.
This transformation will add the following statement to the local
context:
\begin{lstlisting}
search_app_Z : forall (x: Z) (l1 l2: list Z), search x (l1 ++ l2) = (search x l1) || (search x l2)
\end{lstlisting}

There are various ways to handle polymorphism\cite{DBLP:journals/corr/BlanchetteB0S16}\cite{DBLP:conf/frocos/BobotP11}. Among them we chose a monomorphization based on instantiating
the polymorphic types with chosen ground types from the context.

To write the monomorphization tactic\footnote{See file \lstinline 
!elimination_polymorphism.v!.}, the meta-programming language \Ltac\ was our main tool.
Indeed, we did not need a detailed access to the syntax of the terms as \MetaCoq\ provides, 
but we wanted to apply the same tactic to all the hypotheses in a given context (or to a list of polymorphic lemmas). 
\Ltac\ allows matching on the local context in a rather simple way, thanks to the tactic \lstinline !match goal!. 
In details, the instantiation tactic tries to match all hypotheses in the
local context whose type $P$ is a quantified hypothesis: $\forall A \: : \: Type, P'$.
\MetaCoq is useful here, to check that $A$ has type \lstinline!Type!.


Then, the monomorphization tactic scans the goal and instantiates the variable of type \lstinline!Type! with all the subterms of type \lstinline!Type! in the goal. 
All the generated hypotheses are automatically proven by the tactic \lstinline !specialize!. 

In order to avoid infinite loops, the instantiated hypothesis is not added in the context if it is already present. 
The tactic can also take parameters (polymorphic lemmas), and they are monomorphized in the same way. 


In the future, we will run benchmarks to measure the performance of our tactic. 
This may help us to develop heuristics for choosing the instances of lemmas efficiently.

\subsection{Interpreting Algebraic types}
\label{sec:transfos:datatypes}

\newcommand{\pierretac}{{\lstinline!interp_alg_types!}\xspace}
Algebraic datatypes are a special case of inductive types which do not
use non-prenex polymorphism or type dependencies. The epitomy of such
a type is perhaps \lstinline!list! that we used in our examples:
\begin{lstlisting}
Inductive list (A : Type) : Type :=
  | [] : list A
  | cons : A -> list A -> list A.
\end{lstlisting}
where the constructor \lstinline!cons! has the infix notation \lstinline!::!. When one is familiar with inductive types, one knows that this declaration specifies how equality works on the type \lstinline!list!. For instance, \lstinline!x1 :: l1 = x2 :: l2! implies \lstinline!x1 = x2! and \lstinline!l1 = l2!. Moreover, \lstinline![] <> x :: l1!. In general,
in an algebraic datatype \lstinline!I! as defined in \Coq:
\begin{itemize}
\item Each constructor \lstinline!C! of \lstinline!I! is injective, that is:
 \begin{center}
\begin{lstlisting}
forall (x1 y1: A1) ... (xn yn: An),
  C x1 ... xn = C y1 ... yn -> x1 = y1 /\ ... /\ xn = yn
\end{lstlisting}
\end{center}
\item If \lstinline!C! and \lstinline!C'! are two distinct constructors of \lstinline!I!, then their direct images are disjoint, that is:
\begin{center}
\begin{lstlisting}
forall (x1 : A1) ... (xn : An) (x1': A1') ... (xp' : Ap'),
  C x1 ... xn <> C' x1' ... xp'
\end{lstlisting}
\end{center}
Any inhabitant of \lstinline!I! is obtained from one of the \lstinline!C$_{i}$! of \lstinline!I!, that is: 
\begin{lstlisting}
forall (x: I), 
((exists x$_{1,1}$ : A$_{1,1}$)$\ldots$(exists x$_{\ttk_1,1}$ : A$_{1,\ttk_1}$), x = C$_1$ x$_{1,1}\ldots$ x$_{1,\ttk_1}$ )
\/$\ldots$\/((exists x$_{\ttn,1}$ : A$_{\ttn,1}$)$\ldots$(exists x$_{\ttn,\ttk_\ttn}$ : A$_{\ttn,\ttk_\ttn}$), x = C$_\ttn$ x$_{\ttn,1}\ldots$ x$_{\ttn,\ttk_\ttn}$ )                          
\end{lstlisting}
\end{itemize}
Note that the above propositions are statements of first-order logic and
as such, may be communicated to a first-order automatic theorem prover.
However, the third property is written with existential quantifiers
which are not treated by the back-end we use in our proof of concept
(see Sec~\ref{sec:poc}), so our tactic does not generate this property
for the moment.

We define the tactic \pierretac\footnote{See file \lstinline 
!interpretation_algebraic_types.v!.} that finds the algebraic datatypes of \Coq (\eg \lstinline!list Z!) which occur in the goal and automatically proves that (1) their constructors are injective (2) the direct images of their constructors are disjoint.

If we come back to Example~\ref{ex:datatypes}, the datatypes
\lstinline!list! and \lstinline!option! are both made explicit so that
the automatic prover is able to conclude.


\section{Proof of concept}
\label{sec:poc}

As explained in \Sec~\ref{sec:methodology} and the left part of
Figure~\ref{fig:snipe}, the methodology is to combine such
transformations, possibly in different ways, then call an automatic
solver.

In this section, we provide a proof of concept: all the transformations
in the previous section are combined in a fully
automatized tactic called \lstinline!snipe! which applies them and 
sends the resulting goal and 
context to the SMT solver \veriT through the \SMTCoq plugin. This is
illustrated by the right part of Figure~\ref{fig:snipe}.
We now detail this combination and come back to examples 
illustrating the \lstinline!snipe! tactic.

\subsection{Proof strategy}
\label{sec:poc:strategy}

\begin{figure}[!ht]
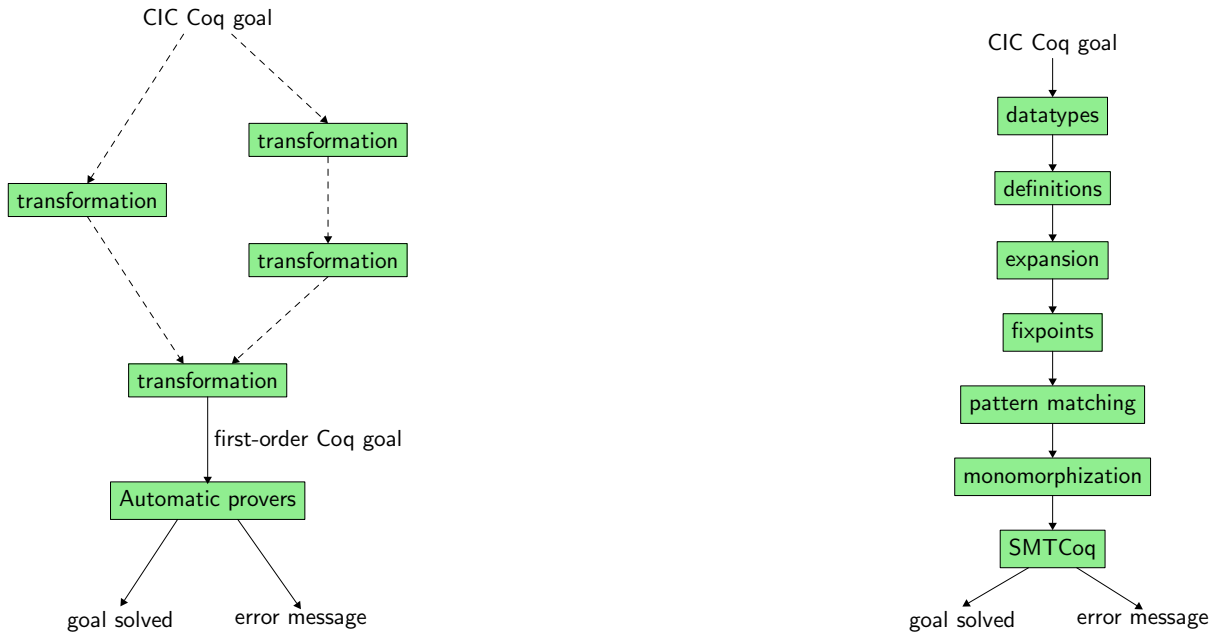

\begin{center}
  \includegraphics[scale=0.8]{figs/hideatp.mps}
  \hfill
  \includegraphics[scale=0.8]{figs/snipe.mps}
  \end{center}
\caption{General methodology (left) and proof of concept (right)}
\label{fig:snipe}
\end{figure}

As presented in Figure~\ref{fig:snipe} we proceed as follows. 
We first 
apply a combination of our transformations in a unique tactic called 
\lstinline !scope!. 
Thanks to the \SMTCoq plugin, we send the goal and the local context 
with additional lemmas obtained by 
the \lstinline!scope! tactic to the external solver \veriT.
Let us describe this two-part process more precisely:

\begin{itemize}
\item \lstinline!scope!: This tactic consists of applying first \pierretac to all algebraic types in the goal and in the context, except types 
already interpreted by the SMT solver like $\mathbb{Z}$ or the Booleans. 
Then it calls the \lstinline !get_definitions! tactic: it adds new definitional hypotheses in the local context, except for the symbols which are part of the built-in theories of \veriT. 
For instance, the definition of the addition in $\mathbb{Z}$ is not needed. 
Then, the tactics \lstinline !expand!, \lstinline!eliminate_fix! and \lstinline !elimination_pattern_matching! are applied to the new generated hypotheses. 
Finally, the monomorphization tactic will assert and prove a new proposition for every polymorphic hypothesis applied to a subterm of type \lstinline!Type! in the goal. 
A tuple of Coq lemmas choosen by the user can be added as parameters to \lstinline !snipe!: the tactic will also try to instantiate them if they are polymorphic. 

\item Once \lstinline !scope! is applied, the tactic \lstinline!verit! is called.
It is an \SMTCoq tactic which solves first-order goals in a combination
of built-in theories (such as linear arithmetic or congruence) by
calling the external SMT solver \veriT and reconstructing a \Coq proof.
Note that:
\begin{itemize}
\item this tactic requires equalities to be decidable (this is the
  proposition \lstinline!CompDec A! mentionned in the introduction)
\item its implementation relies on the \Coq standard library for machine integers and
  arrays, which axiomatize these data-structures; that is why the tactic
  \lstinline!snipe! also relies on these axioms.
\end{itemize}
\end{itemize}

\subsection{Examples}
\label{sec:poc:examples}

Let us go back to our examples\footnote{See file
  \lstinline!examples.v!.}. Example~\ref{ex:datatypes} is
automatically solved with \lstinline!snipe!. Note however that we need
an additional hypothesis
\lstinline!Compdec A!: as previously mentionned the type $A$ has to have
a decidable equality for \SMTCoq to reason on it. The \lstinline!scope!
part of the tactic adds all the hypotheses about the constructors of
types \lstinline!option! and \lstinline!list!, and they are instantiated
by the variable $A$. The definition of \lstinline!hd_error! is also
added in the local context. Since it contains a pattern matching, a proposition
for every pattern is created and proved. The tactic \lstinline!verit!
transforms the first-order hypotheses into assertions for \veriT and
solves the goal.

Example~\ref{ex:search-lemma}, about \lstinline!search!, requires the
instantiation of the previous lemma \lstinline!search_app!. Again, it is solved
automatically by the tactic \lstinline!snipe!, taking this time the
lemma \lstinline!search_app! as a parameter. The proof becomes:
\begin{lstlisting}
Goal forall (A : Type) (H : CompDec A) (x: A) (l1 l2 l3: list A),
  search x (l1 ++ l2 ++ l3) = search x (l3 ++ l2 ++ l1).
Proof. intros A H. snipe @search_app. Qed.
\end{lstlisting}
The instantiation of the polymorphic lemma is done by the monomorphization tactic within \lstinline!scope!.
The part of the proof which required a lemma about Booleans and an
adequate instantiation of it is automatically handled by the
capabilities of \veriT to perform propositional reasoning with quantifier
instantiation.

More interestingly, we can also prove the intermediate lemma in a very
satisfactory way. This lemma is proved by induction on the first list,
and induction is currently out of the scope of most automated provers.
However, once this induction is done, the user should not worry about
proving the sub-cases. This is made possible by our tactic:
\begin{lstlisting}
Lemma search_app : forall {A:Type} {H:CompDec A} x (l1 l2:list A),
   search x (l1 ++ l2) = (search x l1) || (search x l2).
Proof. intros A H x l1 l2. induction l1 as [ | x0 l0 IH]; simpl; snipe. Qed.
\end{lstlisting}

\section{State of the art}
\label{sec:soa}

\paragraph{General automation in proof assistants}

Improving automation in proof assistants based on Type Theory is a
long-standing research topic.

Decision procedures for dedicated theories have met a large success,
witnesses the daily used
\lstinline!ring!~\cite{DBLP:conf/tphol/GregoireM05} and
\lstinline!lia!~\cite{DBLP:conf/types/Besson06} tactics in \Coq, 
deciding equalities respectively in (semi-)ring structures and propositions in the
linear integer arithmetic theory. Automatic tactics also exist for
propositional and first-order logic, such as \lstinline!intuition!,
\lstinline!firstorder! or
\lstinline!crush!\cite{DBLP:books/daglib/0035083} in \Coq. The
limitations of these tactics are that they are useless as soon as the
reasoning requires handling multiple aspects at one time 
(such as propositional logic, arithmetic, equalities...), which is very
often the case when using an interactive theorem prover.

This is why research in this area moved towards making use of more complex
automatic solvers, mainly SMT solvers, which can combine theories with
tableau or first-order provers that usually do not natively handle
theories but better deal with quantifiers. In this direction, two
approaches are usually considered: the {\em autarkic} approach, which
consists in implementing and proving correct the automatic prover in the
proof assistant (\eg the SMT solver \ergo~\cite{DBLP:phd/hal/Lescuyer11}
in \Coq, the tableau prover
\blast~\cite{DBLP:journals/jucs/Paulson99} in Isabelle/HOL, or the
first-order solver \metis in the HOL family~\cite{Hurd05}), and the {\em
  skeptical} approach, which consists in using external provers that
output explanations and only checking these explanations at each execution
(\eg \SMTCoq~\cite{DBLP:conf/cav/EkiciMTKKRB17} in \Coq or
\smt~\cite{DBLP:conf/itp/BohmeW10} in Isabelle/HOL).

However, as we explained, goals in proof assistant usually do not belong
to first-order logic, which is the logic handled by these provers. This
is why encodings need to be performed. In this direction, the most
successful tool is \sledgehammer~\cite{DBLP:conf/lpar/PaulsonB10} for
Isabelle/HOL, which mixes both an autarkic and skeptical approches: it
encodes a higher-order goal, calls many external solvers in parallel
(using lemmas from the global context selected by machine learning), and
uses their answers to reconstruct a proof of the original goal, by
mixing standard tactics with \smt or \metis. This approach was ported to
\Coq in the \coqhammer~\cite{DBLP:journals/jar/CzajkaK18} tool, but with
less success.

We identified one main limitation of \coqhammer to be proof
reconstruction: the tool tries to build a proof of the original goal,
but the external solvers proved the encoded goal. This encoded goal is
far from the original goal because \coqhammer uses a one pass, very
complex encoding of CIC into FOL. This was not a problem for
Isabelle/HOL whose logic is simpler. One cannot rely on the reconstruction of a proof of
the encoded goal, because then it should be proved correct with respect
to the original goal, which is very difficult again because of the
complexity of the encoding.

This is why we proposed this new approach, where the encoding is a
combination of small transformations that can either be certified or output
\Coq proofs (certifying). Then the resulting goal can be checked by any
approach for first-order proving since we do not need to reconstruct
a proof of the original goal. This approach with small, independent
transformations was inspired by other tools that use external automatic
solvers, in particular
\whythree~\cite{DBLP:conf/esop/FilliatreP13}.

As we explained, we are independent of the back-end used to discharge
the first-order goal produced by the transformations. We chose \SMTCoq
in our proof of concept. It restricts us to hypotheses and goals with
only universal and prenex quantification, but offers built-in theories,
which seems a good trade-off for the kind of goals that are commonly
present in \Coq. Targeting first-order provers could also be done by
providing other strategies that would do less work on quantifiers but
encode theories such as linear arithmetic.

We leave for future work a detailed comparison with \coqhammer, both in
terms of performance and expressivity. For this latter, we are currently
theoretically less expressive than \coqhammer (since we only handle a
small part of \Coq logic beyond FOL), but
\begin{itemize}
\item we can detail the fragment of \Coq logic that we handle, whereas
  the success of \coqhammer is more unpredictable,
\item we believe that the approach will scale when implementing more
  involved transformations.
\end{itemize}

Recently, automatic provers have pushed towards more expressivity than
FOL~\cite{DBLP:conf/cade/BarbosaROTB19}. Once they can be used in proof
assistants such as \Coq (currently they do not output certificates for
higher-order aspects), it will be very easy to integrate them in our
setting: it simply requires unplugging the transformations that deal
with the aspects newly handled by these provers. We could also consider
using the theory of algebraic datatypes supported by some SMT solvers to
handle a sub-part of Coq inductive types.

\paragraph{Meta-programming in \Coq}

The approach by certified/certifying transformations relies on
meta-programming. As explained in \Sec~\ref{sec:methodology}, we use two
meta-programming tools available in \Coq: \Ltac and \MetaCoq. These
tools are complementary: \Ltac allows us to handle surface
meta-programming very easily, whereas \MetaCoq allows us to go deeper
into the structure of terms, at the cost of difficult aspects such as De
Bruijn indices. We plan for future work to look at other tools that
could enjoy both worlds, in particular \ltactwo~\cite{pedrot2019ltac2}
and \coqelpi~\cite{DBLP:conf/itp/Tassi19}.

\section{Conclusion and perspectives}
\label{sec:conclusion}

We have developed a methodology to encode some aspects of CIC into FOL by writing 
independent and modular transformations. This modularity is useful to delimit the features of CIC 
we can translate, and to combine the transformations in a chosen order. 
As a proof of concept, we created a \Coq tactic \snipe which performs the transformations described above
and calls an external SMT solver. Some \Coq proofs are now totally automatized thanks to our tactic.

Now that we have the crucial and basic transformations to extend the local 
context with first-order hypotheses, we plan to tackle more complex transformations. 
This future work will help to automatize other aspects of \Coq logic.
Here is a non exhaustive list of transformations or features we would like to treat:
\begin{itemize}
\item \textbf{Encoding of higher-order terms.} We may treat them as usual first-order terms and add an 
applicative symbol $@$ to the language signature. Thus, $f(x)$ becomes 
$@(f, x)$.
\item \textbf{Encoding of inductive predicates.}
They are a particular case of dependent typing, and we would like to obtain first-order statements from them as 
they are very common whenever program specifications are written in \Coq.
\item \textbf{Skolemization.} As we may use external solvers which do not deal with existential quantifiers
after applying our tactic \lstinline!scope!, 
we would like to be able to encode them and thus perform a skolemization on the goal and the hypotheses. 
\end{itemize}

Another important part of our future work is to improve performance and do a benchmark analysis.
This benchmark has to be \emph{qualitative} and \emph{quantitative},
that is, we need to evaluate the efficiency of our tactic
and to know how many goals it can solve. In particular, we will compare \snipe with \coqhammer. 
As previously said, we hope that a more elaborated version of our tactic will do better than \coqhammer 
in the fragment of CIC we chose to deal with.

\bibliographystyle{eptcs}
\bibliography{generic}
\end{document}

%% file: macros.tex
\newcommand{\ignore}[1]{}

\newcommand{\Sec}{\S} 

\newcommand{\ie}{\textit{i.e.}, }
\newcommand{\eg}{\textit{e.g.}, }
\newcommand{\vs}{\textit{vs.} }

\newcommand{\namefont}[1]{\textsf{#1}}

\newcommand{\MetaCoq}{\namefont{MetaCoq}\xspace}
\newcommand{\SMTCoq}{\namefont{SMTCoq}\xspace}
\newcommand{\Ltac}{\namefont{Ltac}\xspace}
\newcommand{\veriT}{\namefont{veriT}\xspace}
\newcommand{\Coq}{\namefont{Coq}\xspace}
\newcommand{\OCaml}{\namefont{OCaml}\xspace}
\newcommand{\snipe}{\texttt{snipe}\xspace}

\newcommand{\ergo}{\texttt{ergo}\xspace}
\newcommand{\blast}{\texttt{blast}\xspace}
\newcommand{\metis}{\texttt{metis}\xspace}
\newcommand{\sledgehammer}{\texttt{sledgehammer}\xspace}
\newcommand{\smt}{\texttt{smt}\xspace}
\newcommand{\coqhammer}{\texttt{CoqHammer}\xspace}
\newcommand{\whythree}{\texttt{Why3}\xspace}
\newcommand{\ltactwo}{\texttt{Ltac2}\xspace}
\newcommand{\coqelpi}{\texttt{Coq-Elpi}\xspace}

\newcommand{\ttk}{\mathtt{k}}
\newcommand{\ttn}{\mathtt{n}}

\newcommand{\ttG}{\mathtt{G}}
\newcommand{\ttP}{\mathtt{P}}